\documentclass[aps,prx,twocolumn,superscriptaddress,showpacs,longbibliography,floatfix,showkeys]{revtex4-1}

\usepackage{amsmath}
\usepackage{graphicx}
\usepackage{color}
\usepackage{hyperref}
\usepackage{float}

\begin{document}

\title{Spatial control of carrier capture in two-dimensional materials:\\ Beyond energy selection rules}

\author{Roberto Rosati}
\affiliation{
Institut f\"ur Festk\"orpertheorie, Universit\"at M\"unster, Wilhelm-Klemm-Str. 10, 48149 M\"unster, Germany}

\author{Frank Lengers}
\affiliation{
Institut f\"ur Festk\"orpertheorie, Universit\"at M\"unster, Wilhelm-Klemm-Str. 10, 48149 M\"unster, Germany}

\author{Doris E. Reiter}
\affiliation{
Institut f\"ur Festk\"orpertheorie, Universit\"at M\"unster, Wilhelm-Klemm-Str. 10, 48149 M\"unster, Germany}
\affiliation{Center for Multiscale Theory and Computation (CMTC), Universit\"at M\"unster,
48149 M\"unster, Germany}

\author{Tilmann Kuhn}
\affiliation{
Institut f\"ur Festk\"orpertheorie, Universit\"at M\"unster, Wilhelm-Klemm-Str. 10, 48149 M\"unster, Germany}

\affiliation{Center for Multiscale Theory and Computation (CMTC), Universit\"at M\"unster,
48149 M\"unster, Germany}


\begin{abstract}
Transition metal dichalcogenide monolayers have attracted wide attention due to their remarkable optical, electronic and mechanical properties. In these materials local strain distributions effectively form quasi zero-dimensional potentials, whose localized states may be populated by carrier capture from the continuum states.
Using a recently developed Lindblad single-particle approach, here we study the phonon-induced carrier capture in a MoSe$_2$ monolayer. Although  one decisive control parameter is the energy selection rule, which links the energy of the incoming carriers to that of the final state via the emitted phonon, we show that additionally the spatio-temporal dynamics plays a crucial role. By varying the direction of the incoming carriers with respect to the orientation of the localized potential, we introduce a new control mechanism for the carrier capture: the spatial control.
\end{abstract}


\maketitle

\section{Introduction}

Monolayers of a transition metal dichalcogenide (TMDC) constitute a new class
of 2D materials which can be direct band gap semiconductors even if their bi-
or multilayer counterparts can have an indirect gap
\cite{Mak10,Splendiani10,Wang12,Kormanyos15}. This makes these material
systems attractive candidates for optical and optoelectronic applications. In
the TMDC monolayer localized quasi zero-dimensional (0D) confinement
potentials can be effectively formed by strain tuning
\cite{Branny16,Krustok16}, resulting in the formation of single photon
emitters \cite{Koperski15,Kumar15,Tonndorf15,Kern16,Branny17}. To populate
such a TMDC quantum dot (QD) the phonon-induced carrier capture from the
extended 2D monolayer states into the 0D QD states could be exploited. Since
controlling and manipulating carriers in 0D structures is a fundamental
requirement for applications, e.g., in the field of quantum information
processing, the carrier capture has been studied in several semiconductor
devices (with both two-
\cite{Ferreira99,Magnusdottir03,Markus04,Nielsen04,Seebeck05,Trumm05,Mielnik15}
or one-dimensional \cite{Schedelbeck97,Lienau00,Wegscheider97} embedding
materials) and by modeling on various theoretical levels, ranging from rate
equations to fully quantum kinetic treatments
\cite{Brum86,Kuhn89,Preisel94,Magnusdottir03,Nielsen04,Reiter09,b-Ferreira15,Lengers17}.
Here we will focus on the capture from 2D into 0D states in a TMDC monolayer.

For an efficient carrier capture, having in mind Fermi's golden rule for the
scattering rates, one might first think of energy selection rules. For
example, if the carrier capture takes place by emission of a phonon, for an
efficient capture the excess energy of the carriers in the 2D system should
be one phonon energy above the energy of the discrete state in the QD. The
capture rate should then be proportional to the squared transition matrix
element between the delocalized initial state and the localized final state.
But this simple picture neglects crucial aspects of the carrier capture on
the nanometer scale. First and foremost, since the electron-phonon
interaction is a local interaction, the carrier capture happens
\emph{locally}, i.e., it should take place only when the carriers are close
to QD. Using a simple rate between the delocalized and localized states would
instantaneously reduce the density in the whole 2D material and not only
close to the QD. This locality is well reproduced by approaches which fully
take into account the off-diagonal nature of the electron density matrix
\cite{Reiter06,Reiter07a,Rosati17}. The local nature, on the other hand,
opens up the possibility to control the capture process beyond the energy
selection rules. Employing a recently established Lindblad single-particle
(LSP) approach \cite{Rosati17} for the electron density matrix, which
combines the ability to correctly treat the locality of scattering processes
and the presence of quantum superposition states with a high numerical
efficiency, we will show that by manipulating only the spatial configuration
of the initial structure, without modifying its energetic characteristics,
the spatio-temporal dynamics of the capture process can be modified in a wide
range.

To be specific, we study the dynamics of an electronic wave packet traveling
in a TMDC monolayer, which impinges on a localized potential forming an
asymmetric QD. We particularly focus on the spatially resolved dynamics in
contrast to the relaxation dynamics of the state population. We show that the
capture process into the localized states of the QD depends sensitively on
the geometry of the problem, in particular, when the QD is rotated with
respect to the wave front of the incoming packet. This opens up the
possibility for a \emph{spatial control}. We demonstrate  spatial control of
the occupations of the QD as well as of the coherences between the discrete
levels of the QD, which result in spatial oscillations of the captured charge
density. Our results highlight the importance of spatial information for
designing future devices using localized potentials in TMDC monolayers.

\section{Theoretical Background}
\subsection{System set-up}\label{sec:Initial}
\begin{figure}[t]
	\centering
	\includegraphics[width=\linewidth]{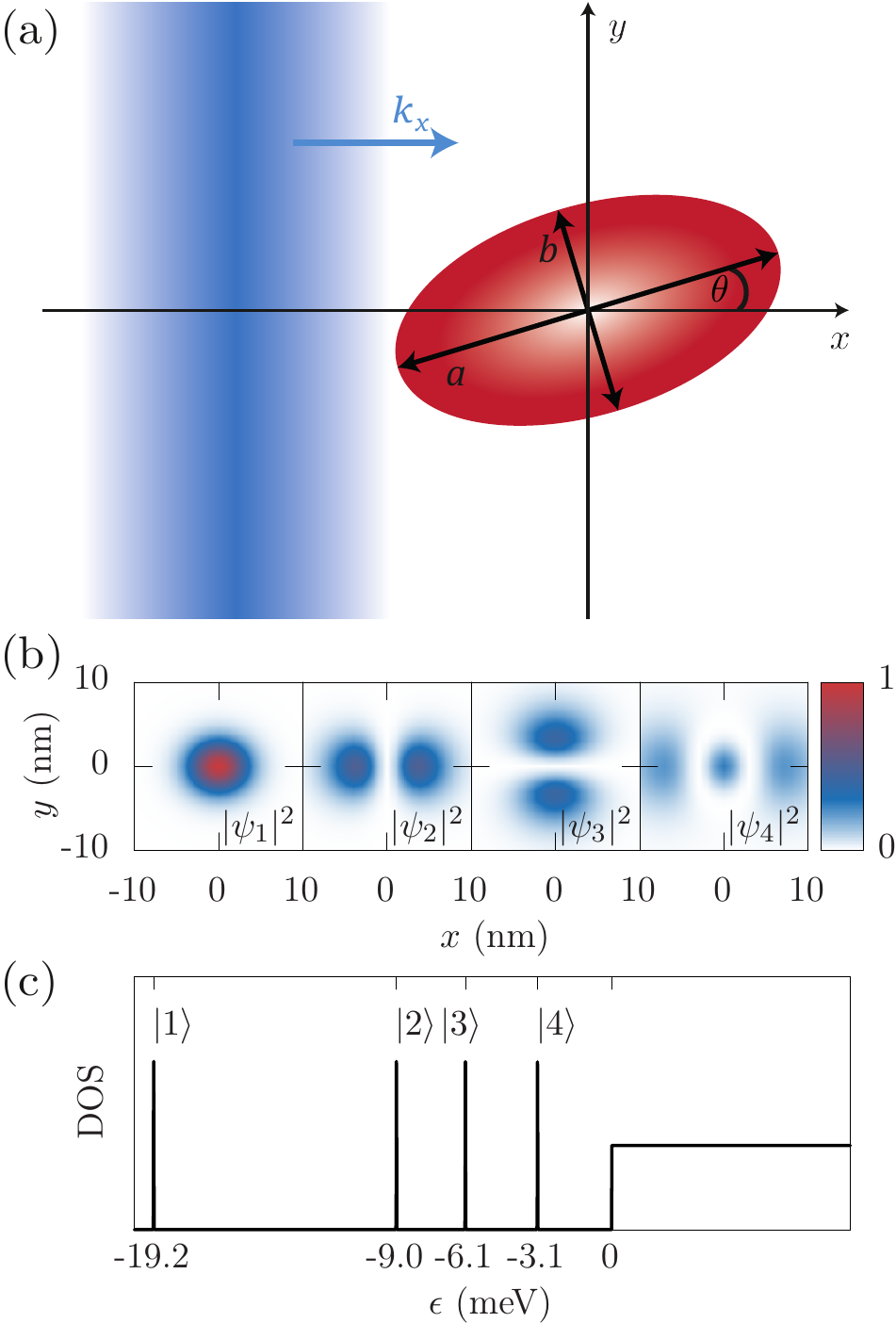}
	\caption{(a) Sketch of the wave packet
impinging on an asymmetric TMDC QD. The semiaxes of the QD are given by $a$ (long semiaxis)
and $b$ (short semiaxis) [see Eq. (\ref{V})]. The angle $\theta$ defines the tilt
between the propagation direction of the incoming wave packet and the long axis
of the QD.  (b) Square moduli
of the bound-states wave functions for the potential in Eq. (\ref{V}) with
$\theta=0$ and (c) corresponding density of states
(DOS) of the conduction band showing the energies of the four bound states with
$\epsilon <0$.
	\label{fig:Fig1}}
\end{figure}

For our studies we consider an electronic wave packet traveling in a TMDC
monolayer impinging on a QD potential as sketched in Fig.~\ref{fig:Fig1}(a).
Carriers can be captured inside the potential by the emission of a
longitudinal optical (LO) phonon with energy $E_{\text{LO}}$, as will be
described in Sec. \ref{sec:dyn}.

For the description of the states we will use the envelope function
formalism, in which the states $|\alpha\rangle$ corresponding to the wave
functions $\psi_\alpha(\mathbf{r})$ are obtained by solving the Schr\"odinger
equation
\begin{equation}\label{SchrE}
	[H_{\text{TMDC}}+V(\mathbf{r})]\psi_\alpha(\mathbf{r})= \epsilon_\alpha \psi_\alpha(\mathbf{r})\,,
\end{equation}
where $\mathbf{r}$ is the 2D position vector in the $xy$-plane,
$H_{\text{TMDC}}$ is the Hamiltonian of the TMDC lattice and $V(\mathbf{r})$
the QD potential. For the confinement potential we assume an elliptically
shaped QD with long semiaxis $a$ and short semiaxis $b$, as sketched in
Fig.~\ref{fig:Fig1}(a), which we model by
\begin{equation}\label{V}
V(\mathbf{r})=- V_0 \text{sech}\left( \sqrt{\frac{x'(\theta)^2}{a^2} +
\frac{y'(\theta)^2}{b^2}} \right) \quad ,
\end{equation}
where sech is the hyperbolic secant and
\begin{equation}\label{rotation}
\left(\begin{matrix}
x'(\theta) \\
y'(\theta)
\end{matrix}\right) = \left( \begin{matrix}
\cos(\theta) & \sin(\theta)\\
-\sin(\theta) & \cos(\theta)
\end{matrix}\right)\left(\begin{matrix}
x \\
y
\end{matrix}\right) \quad .
\end{equation}
According to Eq.~\eqref{rotation}, the long axis of the QD is tilted by the
angle $\theta$ with respect to the $x$-axis.

We restrict ourselves here to a single electronic subband. Details on the
TMDC model used in $H_{\text{TMDC}}$ can be found in the
appendix~\ref{app:TMDC}. The solutions of the Schr\"odinger equation are
composed of continuum states with energies above the band gap and delocalized
over the 2D monolayer, and of $n_b$ bound states ($\alpha=1,\dots,n_b$),
which are spatially localized in the QD region and have a discrete energy
spectrum below the TMDC band minimum.

As a material for our simulations we choose MoSe$_2$ with the material
parameters given in appendix~\ref{app:parameter}. For the QD we set $b=3$~nm,
$a=\sqrt{2}b=4.2$~nm and $V_0=35$~meV, while $\theta$ is a variable tilt
angle. With these parameters the QD has $n_b=4$ bound states at energies
$\epsilon_i$ lying $-19.2$,$-9.0$, $-6.1$ and $-3.1$~meV below the bottom of
the conduction band of the 2D material.  The square moduli of the wave
functions of the bound states are depicted in Fig.~\ref{fig:Fig1}(b) for the
case of $\theta=0$, i.e., for a QD elongated along the $x$-direction. State
$\vert 1 \rangle$ is the ground state with even parity and a  weak elongation
along $x$. The excited states $\vert 2 \rangle$ and $\vert 3 \rangle$ have an
odd-parity and are elongated along the long and short axes of the QD,
respectively, which for $\theta=0$ coincide with the $x$- and $y$-directions.
State $\vert 4 \rangle$ has again even parity and is elongated along the long
axis. The corresponding density of states (DOS) of the structure is
schematically shown in Fig.~\ref{fig:Fig1}(c).

The initial wave packet is chosen to be of wave-front type, which in the
basis of the free TMDC states can be written as
\begin{eqnarray}\label{pureState}
\rho^\circ_{\mathbf{k} + \frac{\mathbf{k}'}{2},\mathbf{k} - \frac{\mathbf{k}'}{2}} & \propto &
 e^{-\frac{1}{2}(k'_x \Delta_x)^2} e^{-\imath k'_x x_0} e^{-\left(\frac{(\hbar^2
k_x^2)/(2 m^*) -E_0}{\sqrt{2}\Delta_E}\right)^2} \notag\\
& \times & \theta(k_x) \delta(k'_{y})
\delta(k_{y}) \, ,
\end{eqnarray}
where $\mathbf{k}=(k_x,k_y)$ is a 2D wave vector and $\theta(k_x)$ is the Heaviside step function. The wave packet has a
finite width in space determined by $\Delta_x=10$ nm and in energy given by
$\Delta_E=5$ meV. It is centered at $x_0=-70$~nm, i.e., sufficiently far from
the QD such that initially there is no overlap with the QD. The excess
energy, which determines the velocity of the wave packet, is taken to be
$E_0= 26.8 \text{ meV} \approx (\epsilon_2+\epsilon_3)/2 + E_{\text{LO}}$,
such that from an energetic point of view the capture into the states $\vert
2 \rangle$ and $\vert 3 \rangle$ should be equally probable.

We emphasize that the key parameter in our study will be the relative
orientation of the wave-packet propagation direction with respect to the
orientation of the elongated QD. Because we fix the propagation direction to
the $x$-direction, the relative orientation is quantified by the tilt angle
$\theta$ of the potential as introduced in Eq.~\eqref{V}.

\subsection{Description of the dynamics}\label{sec:dyn}
To describe the dynamics of the wave packet we set up the equation of motion
for the density matrix $\rho_{\alpha \alpha'}$, which in general can be
written as \cite{Rossi02b,Rosati14e,Rosati13b,Rosati14b,Rosati15}
\begin{equation}\label{eqR}
\frac{d\rho_{\alpha \alpha'}}{dt}=\left. \frac{d\rho_{\alpha \alpha'}}{dt}\right|_{\text{free}}
+\left. \frac{d\rho_{\alpha \alpha'}}{dt}\right|_{\text{scat}}\quad ,
\end{equation}
where $d\rho_{\alpha \alpha'}/dt \vert_{\text{free}}=- \imath
(\epsilon_\alpha - \epsilon_{\alpha'})\rho_{\alpha \alpha'}/\hbar$ gives the
scattering-free contributions, while $d\rho_{\alpha \alpha'}/dt
\vert_{\text{scat}}$ describes the scattering.

The initial wave packet [cf. Eq.~\eqref{pureState}] corresponds to an
excitation of only continuum states, from which a charge transfer into the
bound states may take place by scattering mechanisms. In view of the initial
conditions considered here,  we will concentrate on the carrier capture by
emission of intravalley LO optical phonon modes, whose Fr\"ohlich interaction
induces scattering coefficients $g_{\mathbf{q}}$ between states $\vert
\mathbf{k} + \mathbf{q} \rangle$ and $\vert \mathbf{k} \rangle$ in the form
of \cite{Kaasbjerg12,Kaasbjerg14}
\begin{equation}\label{Froh}
g_{\mathbf{q}} \equiv g_q = \frac{g_{\text{Fr}}}{\sqrt{A}} \text{erfc}(q d /2) \quad ,
\end{equation}
with  $\mathbf{q}$  being the phononic wave vector, while erfc($x$) is the
complementary error function and the constants $g_{\text{Fr}}$ and $d$ depend
on the material. Other scattering mechanisms, i.e., with different phonon
modes or Coulomb interaction, are not efficient here and hence disregarded,
see appendix~\ref{app:phonon}.

In the dynamics of Eq. \eqref{eqR} we describe the scattering using the LSP
equation which we recently developed \cite{Rosati17} by tailoring an
alternative Markov approach \cite{Taj09b,Rosati14e}, the latter already used
for spatio-temporal studies in several materials
\cite{Rosati13b,Rosati14b,Rosati15}. In particular, the scattering terms can
be written as
\begin{equation}\label{LSP}
\left. \frac{d \rho_{\alpha \alpha'}}{dt} \right|_{\text{scat}} \!\!\!=\frac{1}{2}\!
\sum_{\bar{\alpha} \bar{\alpha}', \mathbf{q}} \!\!\left( A^{\mathbf{q}}_{\alpha \bar{\alpha}}
A^{\mathbf{q}*}_{\alpha' \bar{\alpha}'} \rho_{\bar{\alpha} \bar{\alpha}'} \!-\!
A^{\mathbf{q}*}_{\bar{\alpha} \alpha} A^{\mathbf{q}}_{\bar{\alpha} \bar{\alpha}'}
 \rho_{\bar{\alpha}' \alpha'}\right) + \text{H.c.}\, ,
\end{equation}
\vspace{-0.5cm}
where
\begin{equation}\label{LSP_A}
A^{\mathbf{q}}_{\alpha \alpha'} = \sqrt{\frac{2 \pi}{\hbar}} g_{\alpha \alpha';
\mathbf{q}}\frac{e^{-\left( \frac{\epsilon_\alpha - \epsilon_{\alpha'}+E_{\text{LO}}}{2
\bar{\epsilon}} \right)^2}}{(2\pi \bar{\epsilon}^2)^\frac{1}{4}}\quad
\end{equation}
and H.c. denotes Hermitian conjugate
while $g_{\alpha \alpha';\mathbf{q}}=\sum_{\mathbf{k}} \langle \alpha \vert
\mathbf{k} \rangle g_{\mathbf{q}} \langle \mathbf{k} + \mathbf{q} \vert
\alpha' \rangle$. We set $\bar{\epsilon}=3.5$ meV ($\bar{\epsilon}\to 0$) for
the transitions into bound (continuum) states. More details on the approach,
including a discussion about its ability to recover most of the features
obtained in a quantum kinetic descriptions and the meaning of
$\bar{\epsilon}$, may be found in Ref. \cite{Rosati17}. We stress however
that the LSP equation is a Markovian treatment \cite{Taj09b,Rosati14e},
therefore it is computationally much lighter than a full quantum kinetic
approach \cite{Reiter06,Reiter07a,Rossi02b}. This computational lightness of
the LSP approach has allowed us to extend previous studies, which have been
mainly focused on effective 1D systems, to fully 2D systems.
Nevertheless, this treatment is able to describe arbitrary spatially
inhomogeneous carrier distributions and, importantly, naturally includes the
possibility that the final state of a scattering process is given by a
quantum-mechanical superposition state.

The spatio-temporal dynamics of electrons in the TMDC monolayer are obtained
by  numerically integrating the equations of motion. We remind that diagonal
elements of the density matrix $f_\alpha=\rho_{\alpha \alpha}$ are the
\textit{populations}, while the off-diagonal ones $\rho_{\alpha \alpha'}$
with $\alpha'\neq \alpha$ are the \textit{coherences}. Outside the QD region
the continuum 2D states are essentially plane waves, and  the full
single-electron density matrix including both diagonal and off-diagonal
elements has to be taken into account for describing spatial inhomogeneities.
The populations may be interpreted as distribution in energy, while the
distribution in space is provided by
\begin{equation}\label{n}
	n (\mathbf{r})=\sum_{\alpha, \alpha'}  \rho_{\alpha \alpha'}\psi^*_{\alpha'}(\mathbf{r})
\psi_{\alpha}(\mathbf{r}) \, ,
\end{equation}
where $\alpha, \alpha'$ run over all the states  (continuum and bound),  thus
providing the whole electronic distribution (i.e., captured or not).  The
spatial distribution of the trapped charge $n_{\text{QD}} (\mathbf{r})$ is
described analogously when restricting $\alpha,\alpha'$ to the $n_b$ bound states.


\section{Results}

\subsection{Capture dynamics}

\begin{figure*}
\centering
\includegraphics[width=\textwidth]{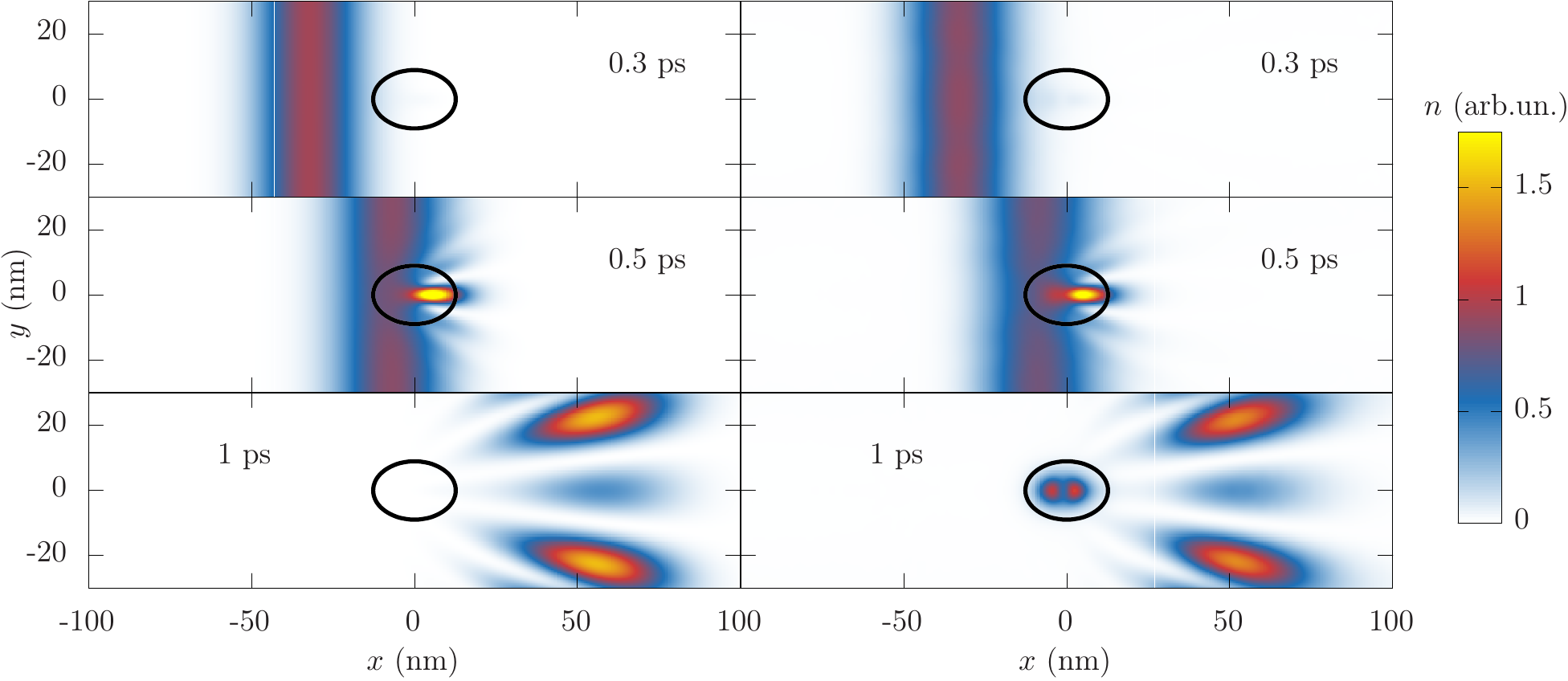}
\caption{Electronic density $n (\mathbf{r})$ for the wave packet
impinging on the QD for $\theta =0$ without (left column) and with (right
column) electron-phonon coupling. The charge has been  normalized to the
height of the initial wave packet. The black line marks the QD region
(defined as where the potential has dropped to $10 \%$ of its maximum). The
three rows show snapshots of the spatio-temporal dynamics for the three
phases: Propagation towards the QD at $t = 0.3$~ps, crossing of the QD at
$t=0.5$~ps and transmission at $t=1.0$~ps. \label{fig:Fig2}}
\end{figure*}

We start our analysis by discussing the spatio-temporal dynamics of the
capture process. Snapshots of the spatio-temporal evolution of the wave
packet without electron-phonon coupling (left column) and with
electron-phonon coupling (right column) are shown in Fig.~\ref{fig:Fig2}. The
dynamics can be separated into three phases: (i) the wave packet propagation
towards the QD (first row), (ii) the crossing of the wave packet over the QD
(second row), and (iii) the motion of the transmitted wave packet (third
row). The wave packet propagation in phase (i) is essentially the same with
and without phonon interaction. At $t=0.3$~ps the wave front has just reached
the QD and the phonon emission is not yet effective. There is only a very
weak redistribution within the continuum states due to the small amount of
occupation above the LO phonon energy. Already the first phase shows that the
locality of the capture is well reproduced by the simulation: Without
overlapping between wave packet and QD the electron-phonon interaction cannot
lead to transitions from continuum into localized states. Technically this
can be traced back to a cancelation between diagonal and off-diagonal
contributions in the equation of motion for the density matrix
\cite{Rosati17}

When arriving at the QD [phase (ii)], the wave front shape of the wave packet
is lost and a  pattern appears. Already on the scattering-free level, we note
an intense peak of charge in the QD region along the $y=0$ line, which is
strongly elongated along $x$.  Although during phase (ii) a capture of
electrons into the localized potential sets in, the density in the QD area is
still rather similar with and without carrier-phonon coupling, showing that
this density is still mainly associated with the continuum states and caused
by their deviations from plane waves above the QD due to their orthogonality
with respect to the bound states. In contrast, after the wave packet has
traversed the QD [phase (iii)], an electronic density remaining in the QD
area is clearly visible only in the presence of electron-phonon interaction.
The fringes of the transmitted wave packet exhibit only slight quantitative
modifications by the scattering processes, while their qualitative shape is
preserved.

\begin{figure}
\centering
\includegraphics[width=\linewidth]{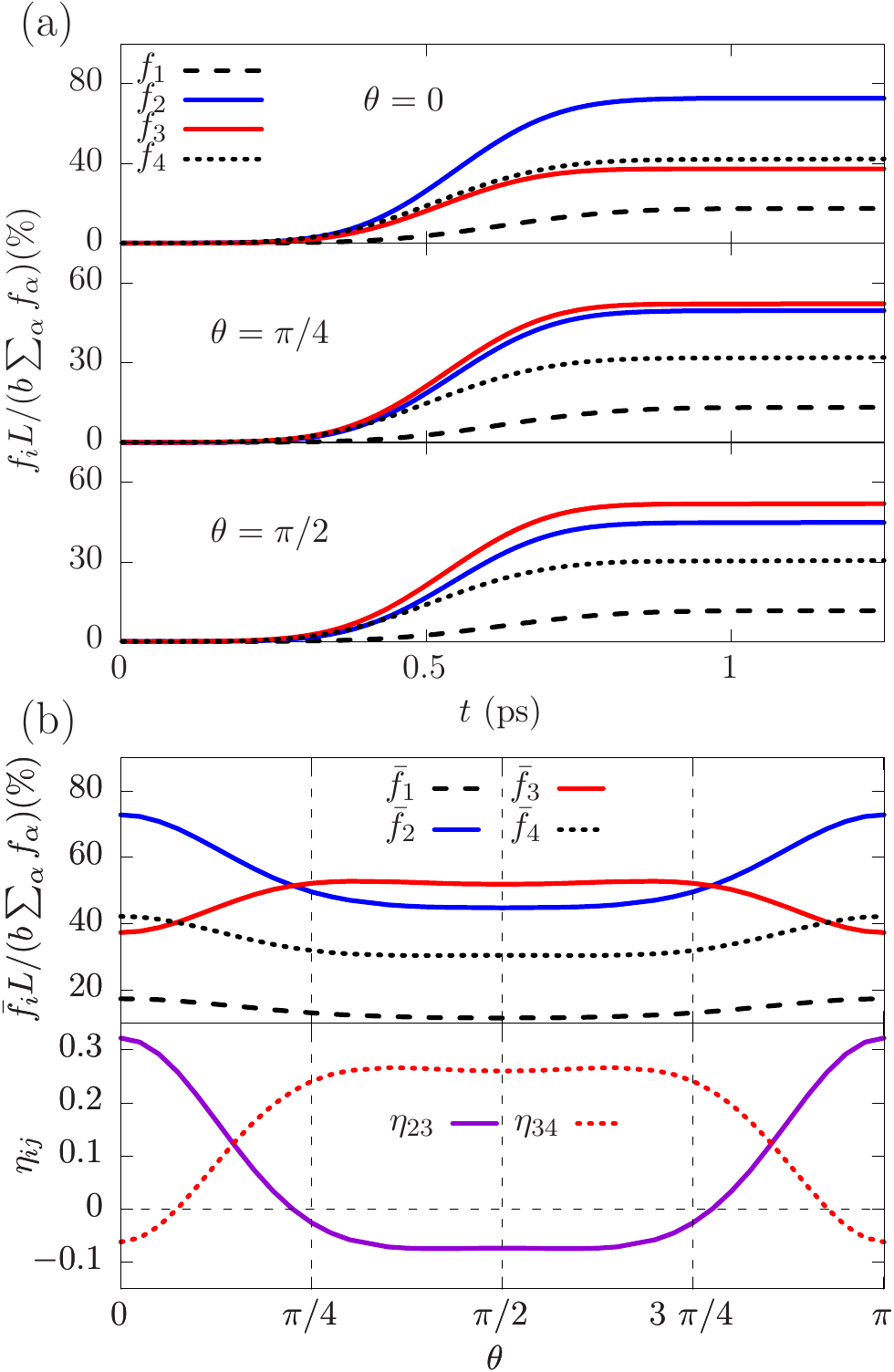}
\caption{(a)
Evolution of the occupations of the bound states for $\theta$=0 (top),
$\theta=\pi/4$ (center) and $\theta=\pi/2$ (bottom). (b) Dependence
of the final occupations $\bar{f}_{i}$ (upper panel) and the relative occupations $\eta_{ij}$ (lower panel)
on the QD orientation $\theta$. All occupations are normalized to the density
contained by the initial electronic distribution in a stripe of height $b/L$.
\label{fig:Fig3}}
\end{figure}

To get a more quantitative picture of the capture dynamics, we now analyze
the occupations $f_i$ of the bound states as a function of time. They are
shown in Fig.~\ref{fig:Fig3}(a) for three different orientations of the QD:
$\theta=0$, $\theta=\pi/4$ and $\theta=\pi/2$. For all occupations we find a
similar behavior: After a certain time, the occupations rise monotonically up
to their respective maximal values and subsequently  stay constant. This
again reflects the locality of the capture process: Only when the wave packet
is in the region of the QD a capture takes place. In analogy with the 1D case
~\cite{Rosati17} we can define a scattering time which is here ranging
between 300 and 400~fs, the exact value depending on the angle $\theta$ and
the state $\vert i \rangle$.

Because the energy distribution of the initial wave packet and the energies
of the bound states are independent of the angle $\theta$, based on energy
selection rules one might expect that the final occupations $\bar{f}_i$
(i.e., the occupations $f_i$ after the wave packet has traversed the QD) are
weakly dependent of the angle $\theta$. In particular, the occupation of the
states $\vert 2 \rangle$ and $\vert 3 \rangle$ should be rather similar due
to the choice of excess energy of the wave packet. However, when looking at
Fig.~\ref{fig:Fig3}(a) we find pronounced differences in the occupations of
these states upon variation of $\theta$. When comparing the final occupations
for a QD elongated along the $x$-direction ($\theta=0$, upper panel) and the
$y$-direction ($\theta=\pi/2$, lower panel), we find that the occupations of
the states $\vert 2 \rangle$ and $\vert 3 \rangle$ are inverted in the two
cases: While for $\theta=0$ the occupation of state $\vert 2 \rangle$ is much
higher than the one of state $\vert 3 \rangle$, for $\theta=\pi/2$ the
occupation of state $\vert 3 \rangle$ is higher than the one of state $\vert
2 \rangle$. Taking into account the spatial shape of the wave functions, we
thus find that capture occurs predominantly into the state which is elongated
along the propagation direction. Indeed, for $\theta=\pi/4$, i.e., when the
QD lies diagonal with respect to the incoming wave packet, the occupations of
$\vert 2 \rangle$ and $\vert 3 \rangle$ are almost the same (central panel).
Remarkably, at $\theta=0$ we find that also $\bar{f}_4$ is bigger than
$\bar{f}_3$, despite its energetic separation from the resonant energy being
much bigger than the one of state $\vert 3\rangle$,  suggesting  a
\textit{spatial selection rule} able to go beyond the pure energetic
considerations.

We quantify the orientation dependence of the captured populations in
Fig.~\ref{fig:Fig3}(b), where in the upper panel we show the final
occupations $\bar{f}_i$ as functions of the angle $\theta$. We find that the
stationary populations corresponding to the states $\vert i \rangle$
elongated along the major axis (i.e., $\bar{f}_1$, $\bar{f}_2$ and
$\bar{f}_4$) decrease by rotating the QD from 0 to $\pi/2$, while the
occupation $\bar{f}_3$ of state $\vert 3 \rangle$, which is elongated along
the orthogonal direction, increases. Thanks to their different spatial
shapes, a rotation of the QD results in the fact that $\bar{f}_3$ overcomes
first $\bar{f}_4$ and then $\bar{f}_2$: This spatial selectivity is
quantified in the lower panel in terms of the relative occupations
\begin{equation}
\eta_{ij}=\frac{\bar{f}_i-\bar{f}_j}{\bar{f}_i+\bar{f}_j} \quad ,
\end{equation}
where the switch in the state occupations is reflected in a change of the
signs.

This $\theta$-dependence is the signature of a \textit{spatial selection
rule} based on the relative orientation of the propagation direction of the
wave packet and the QD long axis, which complements the energy selection rule
determined by the phonon energy $E_{\text{LO}}$.
This implies that
several aspects vary with the angle: the effective scattering matrix
elements, which are determined by the wave packet's propagation direction,
and also the overlap between receiving bound state and emitting traveling wave packet, i.e.,
the interplay between nontrivial spatio-temporal evolution and the locality of the carrier capture.

\subsection{Coherence control}

The spatial control is not limited to the magnitude of the captured
occupations, it further has great impact on the quantum coherences between
the bound states. When the potential consists of several bound states, in
general a capture into a coherent superposition of these states takes place,
resulting in an oscillation of the captured density
\cite{Glanemann05,Reiter06,Rosati17}. Such a build up of superposition states
cannot be described by approaches which treat the capture only in terms of
scattering rates between the different states. However, we have recently
shown that the LSP approach, though being of Markovian nature, indeed
adequately describes this genuine quantum mechanical capture behavior
\cite{Rosati17} because it fully includes off-diagonal elements of the
electron density matrix.

\begin{figure*}
\centering
\includegraphics[width=\textwidth]{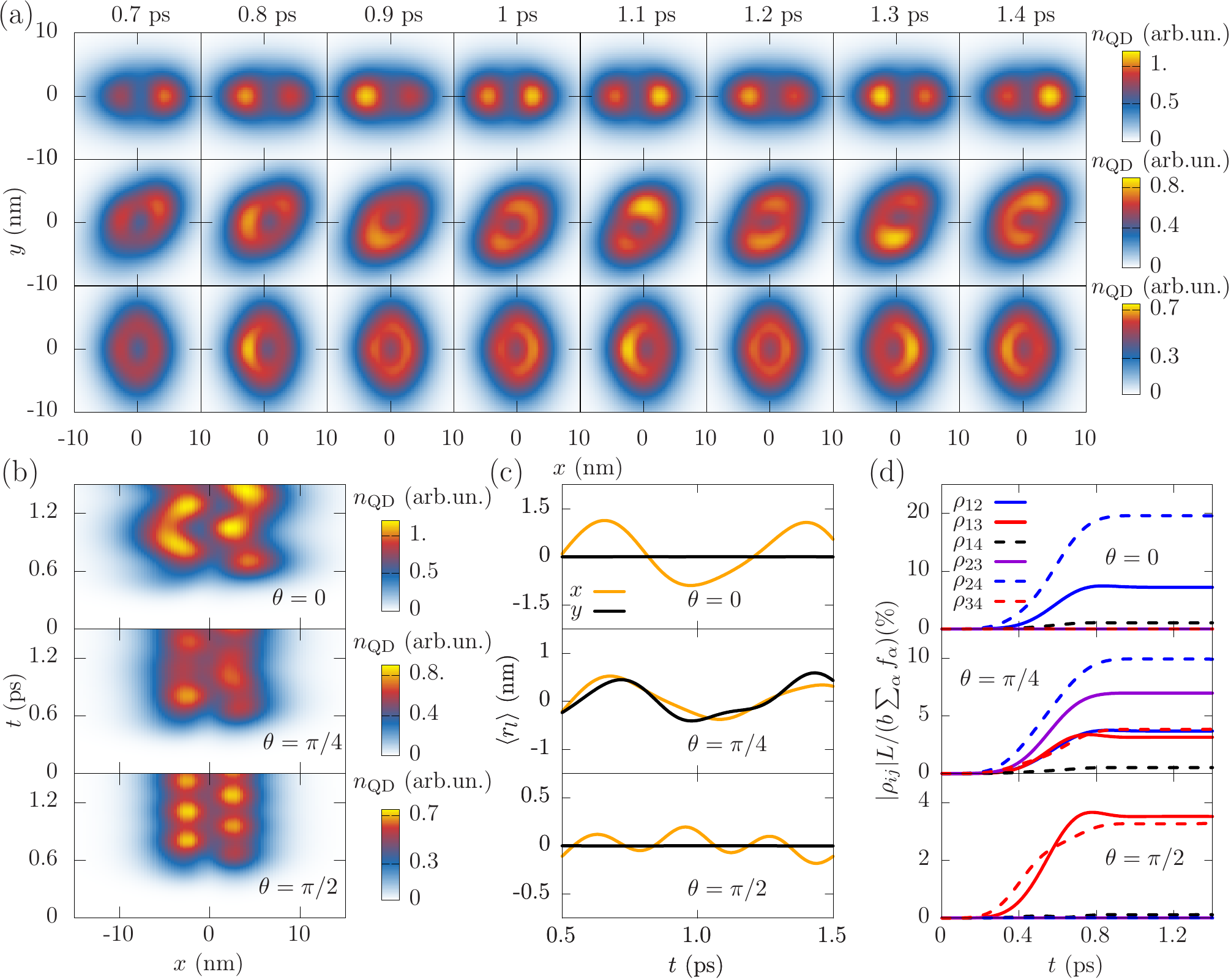}
\caption{(a) Snapshots of the captured charge density
$n_{\text{QD}} (\mathbf{r})$ [see Ancillary Movie 1(a) for the full time evolution]. (b) Spatio-temporal
dynamics of the captured charge density along the $x$-axis [i.e.,
$n_{\text{QD}}(x,y=0,t)$]. (c) Temporal evolution of the $x$-and
$y$-components of the center of mass of the trapped charge distribution [see
also Ancillary Movie 1(b)]. (d) Evolution
of the normalized coherences. All figures are for the three QD orientations
$\theta =0$ (upper row), $\theta = \pi/4$ (central row), and $\theta =
\pi/2$ (lower row). \label{fig:Fig4}}
\end{figure*}

We start analyzing the dynamics of the trapped density $n_{\text{QD}}(x,y)$,
i.e., the density in the subspace of the bound states, for the three
orientations discussed above: Figure~\ref{fig:Fig4}(a) shows snapshots of
this quantity, normalized as in Fig. \ref{fig:Fig2}, while the full time
evolution is displayed in Ancillary Movie 1(a).
Let us start our discussion with the cases of wave-packet propagation along
the long ($\theta=0$, top row) or short ($\theta=\pi/2$, bottom row) QD axis.
In both cases we observe charge oscillations induced by the capture process
along the propagation direction. These oscillations can also be seen in the
top and bottom row of Fig.~\ref{fig:Fig4}(b), where a cut through the
captured density at $y=0$ is plotted as a function of position $x$ and time
$t$. The charge distributions are always symmetric with respect to $y=0$ and
have their maximum at $y=0$. This is also confirmed by looking at the
time-dependence of the center of mass of the charge distribution, defined as
\begin{eqnarray}
\langle r_x \rangle & = & \frac{\int x \, n_{\text{QD}}(x,y) dx dy}{\int n_{\text{QD}}(x,y) dx dy} \ ,\, \notag\\
\langle r_y \rangle & = & \frac{\int y \, n_{\text{QD}}(x,y) dx dy}{\int n_{\text{QD}}(x,y) dx dy}\ ,
\end{eqnarray}
and plotted in Fig.~\ref{fig:Fig4}(c) [see also Ancillary Movie 1(b)]. In the top and bottom panel we find for the transverse
component $\langle r_y\rangle=0$ at any time, while the longitudinal
component $\langle r_x\rangle$ exhibits oscillations which are roughly
sinusoidal. From a closer look we can identify an oscillation period for
$\theta=0$ with $T_{0}\approx 0.7$~ps. For $\theta=\pi/2$ we find the period
to be $T_{\pi/2}\approx0.3$~ps. The time evolutions in space of the center of
mass $(\langle r_x \rangle,\langle r_y \rangle)$ are shown in Ancillary Movie 1(b) (top and bottom row), where we see a
left/right oscillatory motion.

In contrast, for a QD tilted by an angle $\theta=\pi/4$ with respect to the
propagation direction a much more complicated spatio-temporal behavior is
observed [central row in Figs.~\ref{fig:Fig4} and Ancillary Movie 1(a)]. Now we find an oscillation that, although roughly along
the long axis of the QD, is not fixed in time, which can be seen from the
fact that in the central panel of Fig.~\ref{fig:Fig4}(c) the components
$\langle r_x\rangle$ and $\langle r_y\rangle$ are not always proportional to
each other, as it would be the case for a strict oscillation along the long
axis of the QD. In the center of mass motion [see central panel in
Ancillary Movie 1(b)] a quasi-ergodic motion of
$(\langle r_x \rangle, \langle r_y \rangle)$ following the form of the
potential ellipse $V(\mathbf{r})$ is observed. Despite the complexity of the
oscillations, we can still estimate the period to roughly
$T_{\pi/4}\approx0.7$~ps.

The spatio-temporal dynamics of the captured charge density can be understood
when looking at the off-diagonal elements of the density matrix $\rho_{ij}$
in the subspace of the bound states. Their moduli $|\rho_{ij}|$, normalized
in the same way as the diagonal elements in Fig.~\ref{fig:Fig3}, are shown in
Fig.~\ref{fig:Fig4}(d). Like the diagonal elements they show the
characteristic capture behavior: Before the wave packet has reached the QD
region they are zero. Then they build up and  remain constant when the wave
packet has passed the QD. For the coherences the dependence on the
orientation angle $\theta$ is even more pronounced than for the occupations.
For $\theta=0$ (top panel) we find at any time
$\rho_{13}=\rho_{23}=\rho_{34}=0$ while for $\theta=\pi/2$ (bottom panel) we
have $\rho_{12}=\rho_{23}=\rho_{24}=0$. This reflects the symmetry of the
set-up: The incoming wave packet is homogeneous in $y$-direction. Therefore
it can never excite a superposition between a state with even and one with
odd parity in $y$-direction. For $\theta=0$ state $\vert 3 \rangle$ is the
only state with odd parity in $y$-direction. Indeed, no coherences with this
state are excited and the charge distribution keeps its mirror symmetry with
respect to $y=0$. Correspondingly, for $\theta=\pi/2$ state $\vert 2 \rangle$
is the only state with odd parity in $y$-direction and no coherences with
this state are excited. The spatial selection rule is thus able to inhibit
completely the appearance of specific coherences. When looking at the
magnitude of the excited coherences, we find for $\theta=0$ a dominant
excitation of $\rho_{24}$. The corresponding energy difference is
$\epsilon_4-\epsilon_2 = 5.9$~meV giving rise to the oscillation period
$T_{0}\approx0.7$~ps. There is an additional pronounced contribution from
$\rho_{12}$ with an energy difference $\epsilon_2-\epsilon_1 = 10.2$~meV,
from which one can identify a second period $T\approx0.4$~ps which leads to
the fast oscillations in Fig. \ref{fig:Fig4}(b). For $\theta=\pi/2$ there are
two contributions of almost equal strength, $\rho_{13}$ with an energy
difference of $\epsilon_3-\epsilon_1 = 13.1$~meV corresponding to the period
of $T_{\pi/2}\approx0.3$~ps and $\rho_{34}$ with $\epsilon_4-\epsilon_3 =
3.0$~meV, corresponding to a period of about 1.3 ps and which gives rise to
the long time modulations visible in Fig. \ref{fig:Fig4}(c). In both cases
there is a weak excitation of the coherence $\rho_{14}$, which leads to a
slight breathing mode contribution of the charge dynamics. This coherence is
allowed for symmetry reasons but it is strongly suppressed by the weak
overlap of the wave functions and the large energy difference of the states.

If the axis of the QD is tilted by $\theta=\pi/4$ with respect to the
propagation direction of the wave packet, the symmetry selection rules are
relaxed because none of the bound states has a definite parity with respect
to the $y$-axis. Therefore, all the quantum coherences can be excited and, as
can be seen in the central panel of Fig.~\ref{fig:Fig4}(d), they indeed are
all excited. The strongest one is $\rho_{24}$, which gives rise to
oscillations along the long axis with the period $T_{\pi/4}\approx 0.7$~ps,
like in the case $\theta=0$. The next strongest coherence is $\rho_{23}$,
which is neither excited for $\theta=0$ nor for $\theta=\pi/2$. This
coherence induces a rotational-like oscillation in the charge density.
Finally, there are almost equally strong contributions from $\rho_{12}$,
modifying the oscillation along the long axis, as well as from $\rho_{13}$
and $\rho_{34}$, introducing oscillations along the short axis. The
combination of all these coherences gives rise to the complex charge
dynamics.

\section{Conclusion}

In this paper we have shown how in two-dimensional materials the carrier
capture from a traveling electronic wave packet into localized states of an
embedded quantum dot changes with the relative orientation of traveling
direction and quantum dot elongation, despite all the energetic parameters
remain fixed. This proves the effectiveness of spatial selection rules, which
are beyond the usual energetic ones and may find several applications in
controlling charge carrier dynamics on the nanoscale. To be specific, we
considered a monolayer of the transition metal dichalcogenide MoSe$_2$ with a
localization potential as can be formed by a local strain distribution. In
this material, the electrons are efficiently coupled to optical phonons,
which leads to the capture of carriers into the localized states.

To model such a spatial control a theoretical approach is needed which, on
the one hand, fully includes spatially inhomogeneous structures and spatially
inhomogeneous carrier distributions and, on the other hand, is able to
describe genuine quantum features like capture processes into coherent
superposition states and the subsequent dynamics of these superpositions. For
this purpose we have employed a recently developed Lindblad single particle
approach in the density matrix formalism including electron-phonon
scattering. A big advantage of this approach compared to, e.g., a fully
quantum kinetic treatment, is its strongly reduced computational complexity,
which allowed us to simulate the full 2D problem discussed here. Though
treating the interaction processes on a Markovian level, this approach has
been shown to well reproduce  the \emph{locality} of the scattering process,
a basic ingredient to describe the spatial control employed here.

To be specific, we have considered a wave packet traveling in a MoSe$_2$
monolayer impinging on an asymmetric localized potential with bound states. The carriers
can be trapped into the bound states by emission of LO phonons. We have shown
that the spatial control realized by varying the angle between wave-packet
propagation direction and long axis of the quantum dot affects two aspects of the
carrier capture: (a) The occupations of the bound states depend sensitively
on the angle and can be varied significantly by changing the orientation.
This happens despite the energy selection rules do not change when
varying the relative orientation. (b) The coherences between bound states,
which build up during the capture process, strongly depend on the
orientation. Specific coherences can be entirely switched off in the case of
highly symmetric configurations of wave packet and quantum dot orientation.
The capture into superpositions of the bound states is particularly visible
in the spatio-temporal dynamics of the trapped density, which shows an
oscillatory behavior. The period of the oscillations depends on the involved
states and hence is a direct measure for the strength of the coherences. For
a less symmetric situation, e.g., for a tilt angle of $\theta=\pi/4$, a large
number of coherences may be excited by the capture process leading to a
complicated spatio-temporal dynamics.

In conclusion, the locality of carrier capture is crucial for a correct
description of such processes. This has allowed us to exploit a
\emph{spatial} control of carrier capture processes beyond the energy
selection rules. In the process of miniaturization, the spatio-temporal
dynamics will play a more and more decisive role. In this context, our
studies establish the foundations for describing and exploiting the spatial
control of charge carrier dynamics in 2D systems.

\section*{Acknowledgements}

We thank Daniel Wigger for useful discussions.

\appendix

\section{TMDC structure}\label{app:TMDC}
A free-standing monolayer TMDC has a hexagonal lattice with direct band gap
at $\mathbf{K}$ and $\mathbf{K}'$ valleys, where the Hamiltonian results in a
single electron dispersion relation reading $E_{\mathbf{k},\bar b,\bar s,\bar
v} = \bar v~\bar s~\frac{\lambda_c+\lambda_v}{2}~+~a_{0}~t~\bar
b~\sqrt{[(\Delta_G - \bar v\bar s(\lambda_v - \lambda_c))/(2 a_{0} t)]^2 +
|\mathbf{k}|^2}$, where $\mathbf{k}$ is a two-dimensional wave vector, the
constants $\Delta_G$, $\lambda_{c/v}$, providing respectively band gap and
half conduction/valence band splitting, and the parameters $a_{0}$ and $t$
depend on the specific TMDC, while the label $\bar b=\pm$ 1 stands for
conduction/valence band, $\bar s=\pm 1$ for spin up/down and $\bar v=\pm$1
for valley $\mathbf{K}/\mathbf{K}'$  \cite{Shan13,Ferreiros14,Xiao12}.
Although a TMDC monolayer has an involved band structure
\cite{Ding11,Zhu11,Ugeda14,Berghauser14}, in view of the scales involved (see
appendix~\ref{app:phonon}) in this work we restrict ourselves to one subband
with $\bar b=\bar s=\bar v=1$ and to a region close to its minimum, where the
dispersion relation is almost parabolic and the associated eigenstates may be
approximated as scalar states like in conventional semiconductors,
$\psi_{\mathbf{k}}(\mathbf{r})\equiv \langle \mathbf{r} \vert \mathbf{k}
\rangle =  e^{\imath \mathbf{k}\cdot \mathbf{r}}/\sqrt{A} $, with $A=L^2$
being the normalization area of the two-dimensional device.
Note that Eq.~\eqref{SchrE} has been solved ba expanding the wave functions
$\psi_{\alpha}(\mathbf{r})$ in these states $\vert \mathbf{k}\rangle$ with
$\langle \mathbf{k} \vert H_{\text{TMDC}}
\vert\mathbf{k}'\rangle=E_{\mathbf{k},1,1,1}
\delta_{\mathbf{k},\mathbf{k}'}$.

The excess energy $E_0$ we chose lies just above the minimum of the subband,
i.e., where the scalar parabolic approximation works more effectively and the
excitonic effects should be of minor importance
\cite{Klots14,Qu17,Steinleitner17}, in contrast to what happens far below the
band gap, where the excitonic effects should be typically dominant
\cite{Ugeda14,He14,Berghauser14,Chernikov14,Feierabend17}.

\section{Scattering mechanisms} \label{app:phonon}
In view of the energetic separation between continuum and bound states we
disregard the intravalley acoustical phonon modes. In general, TMDCs have six
optical modes, of which however only two -- the so-called LO and $A_1$ modes
-- are able to effectively influence the electron dynamics \cite{Sohier16}.
For MoSe$_2$ the electron-LO phonon coupling coefficients in the long
wavelength limit are one order of magnitude bigger than the electron-$A_1$
phonon ones \cite{Sohier16}; as a consequence, here we restrict ourselves to
intravalley LO phonons with a fixed energy of $E_{\text{LO}} \approx 34.4$~meV
\cite{Sohier16}. We consider the low temperature limit, $k_B T \ll
E_{\text{LO}}$ ($T$ denoting the temperature and $k_B$ Boltzmann's constant),
in which only (spontaneous) phonon emission processes are possible. The
surrounding material \cite{Ugeda14,Yu16} can modify the electron-LO phonon
scattering coefficients \cite{Sohier16}: This would however affect mostly the
quantitative magnitude of the captured charge, and only in a minor way the
qualitative features discussed here. In view of chiral optical selection
rules and spin splittings, a wave packet initially located in the  $\bar s=
\bar v=1$ subband can be realized by circularly polarized excitation
\cite{Xiao12}. Although intervalley scattering mechanisms could in principle
transfer charge from $\textbf{K}$ to $\textbf{K}'$, the intervalley
relaxation time in TMDCs is of several picoseconds in the low temperature
limit \cite{Molina17}. In addition, here the spin preserving intervalley
transitions are strongly suppressed  in view of our excess energy $E_0$ lying
very close to the minimum of the subband with same spin in $\textbf{K}'$
(located 2$|\lambda_C|$=21 meV above the minimum of the subband with $\bar
s=1$ in $\mathbf{K}$). Spin-flipping processes induce slow relaxation times
of the order of tens of picoseconds at low temperatures \cite{Song13}. In
view of the sub-picosecond time scale considered here (see, e.g., Fig.
\ref{fig:Fig3}) we thus restrict our attention to one subband. In this work
we consider low-density excitations, where the Coulomb-induced scattering is
negligible as well \cite{Glanemann05,Rosati17}, such that we do not take the
Coulomb interaction into account in our present studies.

\section{Material parameters} \label{app:parameter}
 In this work we focus on MoSe$_2$, whose above-introduced dispersion relation is given by
material parameters $\lambda_c=-10.5$~meV \cite{Liu13}, $\lambda_v= 90$~meV,
$a_{0}\approx 3.3~$\AA, $\Delta_G=1.47$~eV and $t=0.94$~eV \cite{Xiao12},
resulting in an effective mass of $m^* =0.54 m_0$ for $\bar b=\bar s=\bar
v=1$ ($m_0$ being the free electron mass). Concerning the Fr\"ohlich
interaction of Eq. (\ref{Froh}), the effective layer thickness $d$ and
$g_{\text{Fr}}$ have been given in \cite{Kaasbjerg14} for MoS$_2$;
considering the differences in the material parameters of MoS$_2$ and
MoSe$_2$ \cite{Sohier16}, here we use $d=5.36$ \AA\ and $g_{\text{Fr}}=419.7
\text{meV}$\AA.


\begin{thebibliography}{10}

\bibitem{Mak10}
K. F. Mak, C. Lee, J. Hone, J. Shan and T.~F. Heinz, Atomically thin
  ${\mathrm{MoS}}_{2}$: A new direct-gap semiconductor, Phys. Rev.
  Lett. \textbf{105,} 136805 (2010).

\bibitem{Splendiani10}
A. Splendiani,  L. Sun, Y. Zhang, T. Li, J. Kim, C.-Y. Chim, G. Galli and F. Wang, Emerging photoluminescence in monolayer $\mathrm{MoS_2}$,  Nano
  Lett. \textbf{10,} 1271--1275 (2010).

\bibitem{Wang12}
 Q.~H. Wang, K. Kalantar-Zadeh, A. Kis, J.~N. Coleman and M.~S. Strano,
  Electronics and optoelectronics of two-dimensional transition metal
  dichalcogenides, Nat. Nanotechnol. \textbf{7,} 699--712 (2012).

 \bibitem{Kormanyos15}
A. Korm\'{a}nyos, G. Burkard, M. Gmitra, J. Fabian, V. Z\'olyomi, N. D. Drummond and V. Fal'ko. $\mathrm{k\cdot p}$ theory for two-dimensional transition metal
  dichalcogenide semiconductors, 2D Materials \textbf{2,}
  022001 (2015).

\bibitem{Branny16}
A. Branny, G. Wang, S. Kumar, C. Robert, B. Lassagne, X. Marie, B. D. Gerardot and B. Urbaszek, Discrete quantum dot like emitters in monolayer $\mathrm{MoSe_2}$.
  Spatial mapping, magneto-optics, and charge tuning, Appl. Phys.
  Lett. \textbf{108,} 142101 (2016).

\bibitem{Krustok16}
J. Krustok {\em et al.}, Optical study of local
  strain related disordering in CVD-grown $\mathrm{MoSe_2}$ monolayers, Appl.
  Phys. Lett. \textbf{109,} 253106 (2016).

\bibitem{Koperski15}
M. Koperski.  K. Nogajewski, A. Arora, V. Cherkez, P. Mallet, J.-Y. Veuillen, J. Marcus, P. Kossacki and M. Potemski, Single photon emitters in
  exfoliated $\mathrm{WSe_2}$ structures, Nat. Nanotechnol. \textbf{10,} 503--506 (2015).

\bibitem{Kumar15}
S. Kumar, A. Kaczmarczyk and B.~D. Gerardot, Strain-induced spatial and
  spectral isolation of quantum emitters in mono- and bilayer $\mathrm{WSe_2}$. Nano
  Lett. \textbf{15,} 7567--7573 (2015).

\bibitem{Tonndorf15}
Tonndorf,~P. {\em et al.}, Single-photon emission from localized excitons in an
  atomically thin semiconductor, Optica \textbf{2,} 347--352 (2015).

\bibitem{Kern16}
Kern,~J. {\em et al.}, Nanoscale positioning of single-photon emitters in
  atomically thin $\mathrm{WSe_2}$, Adv. Mater. \textbf{28,} 7101--7105, (2016).


\bibitem{Branny17}
A. Branny, S. Kumar, R. Proux, and B.~D. Gerardot, Deterministic
  strain-induced arrays of quantum emitters in a two-dimensional
  semiconductor, Nat. Commun. \textbf{8,} 15053 (2017).


\bibitem{Ferreira99}
R. Ferreira and G. Bastard, Phonon-assisted capture and intradot Auger
  relaxation in quantum dots, Appl. Phys. Lett. \textbf{74,}
  2818--2820 (1999).

\bibitem{Magnusdottir03}
I. Magnusdottir, S. Bischoff, A.~V. Uskov, and J. M\o{}rk, Geometry
  dependence of Auger carrier capture rates into cone-shaped self-assembled
  quantum dots, Phys. Rev. B \textbf{67,} 205326 (2003).

\bibitem{Markus04}
A. Markus and A. Fiore, Modeling carrier dynamics in quantum-dot lasers,
  Phys. Status Solidi A \textbf{201,} 338--344 (2004).

\bibitem{Nielsen04}
T.~R. Nielsen, P. Gartner and F. Jahnke, Many-body theory of carrier capture
  and relaxation in semiconductor quantum-dot lasers, Phys. Rev. B
  \textbf{69,} 235314 (2004).

\bibitem{Seebeck05}
J. Seebeck, T.~R. Nielsen, P. Gartner and F. Jahnke, Polarons in
  semiconductor quantum dots and their role in the quantum kinetics of carrier
  relaxation, Phys. Rev. B \textbf{71,}  125327 (2005).

\bibitem{Trumm05}
S. Trumm and M. Wesseli, Spin-preserving ultrafast carrier capture and relaxation in InGaAs
  quantum dots, Appl. Phys. Lett. \textbf{87,}  153113 (2005).

\bibitem{Mielnik15}
A. Mielnik-Pyszczorski, K. Gawarecki and P. Machnikowski, Phonon-assisted tunneling of electrons in a quantum well/quantum dot injection structure, Phys. Rev. B \textbf{91,}  195421 (2015).

\bibitem{Schedelbeck97}
G. Schedelbeck, W. Wegscheider, M. Bichler and G. Abstreiter, Coupled
  quantum dots fabricated by cleaved edge overgrowth: From artificial atoms to
  molecules, Science \textbf{278,} 1792--1795 (1997).

\bibitem{Lienau00}
Ch. Lienau, V. Emiliani, T. Guenther, F. Intonti, T. Elsaesser, R. N\"otzel and K.H. Ploog, Near field optical spectroscopy of confined excitons,
  Phys. Status Solidi A \textbf{178,} 471--479 (2000).

\bibitem{Wegscheider97}
W. Wegscheider, G. Schedelbeck, G. Abstreiter, M. Rother and M. Bichler,
  Atomically precise GaAs/AlGaAs quantum dots fabricated by twofold cleaved
  edge overgrowth, Phys. Rev. Lett. \textbf{79,} 1917--1920 (1997).

\bibitem{Brum86}
 J.~A. Brum, and G. Bastard, Resonant carrier capture by semiconductor quantum
  wells, Phys. Rev. B \textbf{33,} 1420--1423 (1986).

\bibitem{Kuhn89}
T. Kuhn and G. Mahler, Carrier capture in quantum wells and its importance
  for ambipolar transport, Solid-state electronics \textbf{32,} 1851--1855 (1989).

\bibitem{Preisel94}
M. Preisel and J. M\o{}rk, Phonon-mediated carrier capture in quantum well
  lasers, J. Appl. Phys. \textbf{76,} 1691--1696 (1994).

\bibitem{Reiter09}
D.~E. Reiter, E.~Y. Sherman, A. Najmaie, and J.~E. Sipe, Coherent control of
  electron propagation and capture in semiconductor heterostructures,
  EPL \textbf{88,} 67005 (2009).

\bibitem{b-Ferreira15}
R. Ferreira and G. Bastard, Capture and Relaxation in Self-Assembled
  Semiconductor Quantum Dots,
2053-2571 (Morgan and Claypool Publishers, San Rafael, 2015).

\bibitem{Lengers17}
F. Lengers, R. Rosati, T. Kuhn, and D.~E. Reiter, Spatio-temporal dynamics of
  carrier capture processes: Simulation of optical signals, Acta Physica
  Polonica A \textbf{132,} 372--375 (2017).

\bibitem{Reiter06}
D. Reiter, M. Glanemann, V.~M. Axt and T. Kuhn, Controlling the capture
  dynamics of traveling wave packets into a quantum dot, Phys. Rev. B
  \textbf{73,} 125334 (2006).

\bibitem{Reiter07a}
D.Reiter, M. Glanemann, V.~M. Axt and T. Kuhn Spatiotemporal dynamics in
  optically excited quantum wire-dot systems: Capture, escape, and wave-front
  dynamics, Phys. Rev. B \textbf{75,} 205327 (2007).

\bibitem{Rosati17}
R. Rosati, D.~E. Reiter and T. Kuhn, Lindblad approach to spatiotemporal
  quantum dynamics of phonon-induced carrier capture processes, Phys.
  Rev. B \textbf{95,} 165302 (2017).


\bibitem{Rosati14e}
R. Rosati, R.~C. Iotti, F. Dolcini and F. Rossi, Derivation of nonlinear
  single-particle equations via many-body lindblad superoperators: A
  density-matrix approach, Phys. Rev. B \textbf{90,} 125140 (2014).

\bibitem{Rosati13b}
R. Rosati and F. Rossi, Microscopic modeling of scattering quantum
  non-locality in semiconductor nanostructures, Appl. Phys. Lett.,
  \textbf{103,} 113105 (2013).

\bibitem{Rosati14b}
R. Rosati and F. Rossi, Scattering nonlocality in quantum charge transport: Application to semiconductor nanostructures, Phys. Rev. B ,
  \textbf{89,} 205415 (2014).

\bibitem{Rosati15}
R. Rosati, F. Dolcini and F. Rossi, Dispersionless propagation of electron
  wavepackets in single-walled carbon nanotubes,  Appl. Phys. Lett.,
  \textbf{106,} 243101 (2015).



\bibitem{Rossi02b}
F. Rossi and T. Kuhn, Theory of ultrafast phenomena in photoexcited
  semiconductors, Rev. Mod. Phys. \textbf{74,} 895--950 (2002).

\bibitem{Kaasbjerg12}
K. Kaasbjerg, K.~S. Thygesen and  K.~W. Jacobsen, Phonon-limited mobility in
  $n$-type single-layer $\mathrm{MoS_2}$ from first principles, Phys. Rev.
  B \textbf{85,} 115317 (2012).


\bibitem{Kaasbjerg14}
K. Kaasbjerg, K.~S. Bhargavi and S.~S. Kubakaddi, Hot-electron cooling by
  acoustic and optical phonons in monolayers of ${\mathrm{MoS}}_{2}$ and other
  transition-metal dichalcogenides, Phys. Rev. B \textbf{90,} 165436 (2014).






\bibitem{Taj09b}
D. Taj, R.~C. Iotti and F. Rossi, Microscopic modeling of energy relaxation
  and decoherence in quantum optoelectronic devices at the nanoscale,
  Eur. Phys. J. B \textbf{72,} 305--322 (2009).

\bibitem{Glanemann05}
M. Glanemann, V.~M. Axt, and T. Kuhn, Transport of a wave packet through
  nanostructures: Quantum kinetics of carrier capture processes, Phys.
  Rev. B \textbf{72,} 045354 (2005).




\bibitem{Shan13}
W.-Y. Shan, H.-Z. Lu and D. Xiao, Spin Hall effect in spin-valley coupled monolayers of transition metal dichalcogenides, Phys.
  Rev. B \textbf{88,} 125301 (2013).

\bibitem{Ferreiros14}
Y. Ferreiros, and A. Cortijo, Large conduction band and Fermi velocity spin
  splitting due to Coulomb interactions in single-layer ${\mathrm{MoS}}_{2}$, Phys. Rev. B \textbf{90,} 195426 (2014).

\bibitem{Xiao12}
D. Xiao, G.-B. Liu, W. Feng, X. Xu and W. Yao, Coupled spin and valley
  physics in monolayers of ${\mathrm{MoS}}_{2}$ and other group-VI
  dichalcogenides, Phys. Rev. Lett. \textbf{108,} 196802 (2012).

\bibitem{Ding11}
Y. Dinga, Y. Wang, J. Ni, L. Shi, S. Shi and W. Tang, First principles study
  of structural, vibrational and electronic properties of graphene-like $\mathrm{MX_2}$
  (M=Mo, Nb, W, Ta; X=S, Se, Te) monolayers, Physica B: Condensed
  Matter \textbf{406,} 2254--2260 (2011).

\bibitem{Zhu11}
Z.~Y. Zhu, Y.~C. Cheng and U. Schwingenschl\"ogl, Giant spin-orbit-induced
  spin splitting in two-dimensional transition-metal dichalcogenide
  semiconductors, Phys. Rev. B \textbf{84,} 153402 (2011).

\bibitem{Ugeda14}
M.~M. Ugeda {\em et al.},
Giant bandgap renormalization and excitonic effects in a monolayer
  transition metal dichalcogenide semiconductor, Nat. Mater.
  \textbf{13,} 1091--1095 (2014).

\bibitem{Berghauser14}
G. Bergh\"auser and E. Malic, Analytical approach to excitonic properties of
  $\mathrm{MoS_2}$, Phys. Rev. B \textbf{89,} 125309 (2014).

 \bibitem{Klots14}
Klots,~A. {\em et al.}, Probing excitonic states in
  suspended two-dimensional semiconductors by photocurrent spectroscopy,
  Scientific reports \textbf{4,} 6608 (2014).

\bibitem{Qu17}
F. Qu, A. Dias, J. Fu, L. Villegas-Lelovsky and D.~L. Azevedo, Tunable spin
  and valley dependent magneto-optical absorption in molybdenum disulfide
  quantum dots, Scientific reports \textbf{7,}  41044 (2017).

\bibitem{Steinleitner17}
P. Steinleitner, P. Merkl, P. Nagler, J. Mornhinweg, C. Schüller, T. Korn, A. Chernikov and R. Huber, Direct observation of ultrafast exciton
  formation in a monolayer of $\mathrm{WSe_2}$, Nano Lett. \textbf{17,} 1455--1460 (2017).

\bibitem{He14}
K. He, N. Kumar, L. Zhao, Z. Wang, K. F. Mak, H. Zhao and J. Shan, Tightly
  bound excitons in monolayer ${\mathrm{WSe}}_{2}$, Phys. Rev. Lett. \textbf{113,} 026803 (2014).

\bibitem{Chernikov14}
A. Chernikov, T. C. Berkelbach, H. M. Hill, A. Rigosi, Y. Li, O. B. Aslan, D- R. Reichman, M. S. Hybertsen and T. F. Heinz, Exciton binding energy
  and nonhydrogenic Rydberg series in monolayer ${\mathrm{WS}}_{2}$,
  Phys. Rev. Lett. \textbf{113,} 076802 (2014).

\bibitem{Feierabend17}
M. Feierabend, G. Bergh{\"a}user, A. Knorr and E. Malic, Proposal for dark
  exciton based chemical sensors, Nat. Commun. \textbf{8,} 14776 (2017).

\bibitem{Sohier16}
T. Sohier, M. Calandra, and F. Mauri, Two-dimensional Fr\"ohlich interaction
  in transition-metal dichalcogenide monolayers: Theoretical modeling and
  first-principles calculations, Phys. Rev. B \textbf{94,} 085415 (2016).



\bibitem{Yu16}
Y. Yu, Y. Yu, C. Xu, Y.-Q. Cai, L. Su, Y. Zhang, Y.-W. Zhang, K. Gundogdu and L. Cao, Engineering substrate interactions for high luminescence efficiency
  of transition-metal dichalcogenide monolayers, Adv. Funct.
  Mater. \textbf{26,} 4733--4739 (2016).

\bibitem{Molina17}
A. Molina-S\'{a}nchez, D. Sangalli, L. Wirtz and A. Marini, Ab initio
  calculations of ultrashort carrier dynamics in two-dimensional materials:
  Valley depolarization in single-layer $\mathrm{WSe_2}$, Nano Lett., \textbf{17,}
4549-4555 (2017).

\bibitem{Song13}
Y. Song and H. Dery, Transport theory of monolayer transition-metal dichalcogenides through symmetry,
  Phys. Rev. Lett. \textbf{111,} 026601 (2013).



\bibitem{Liu13}
 G.-B. Liu, W.-Y. Shan, Y. Yao, W. Yao and D. Xiao, Three-band tight-binding
  model for monolayers of group-VIB transition metal dichalcogenides,
  Phys. Rev. B \textbf{88,} 085433 (2013).





















\end{thebibliography}
\end{document}